\documentclass[12pt]{article}

\usepackage{amsmath}
\usepackage{amssymb}

\advance\hoffset by -7mm   
\makeatletter 
\@addtoreset{equation}{section}  
\makeatother 
 
\def\ii{\'\i}

\def\cao{\c c\~ao}

\def\ftoday{{\sl {Le \number\day \space\ifcase\month 
\or janvier\or f\'evrier\or mars\or avril\or mai
\or juin\or juillet\or ao\^ut\or septembre\or octobre
\or novembre \or d\'ecembre\fi\space \number\year}}}    
\def\ptoday{{\sl {\number\day \space de\space \ifcase\month 
\or janeiro\or fevereiro\or mar{\c c}o\or abril\or maio
\or junho\or julho\or agosto\or setembro\or outubro
\or novembro \or dezembro\fi\space de\space \number\year}}}    
\def\gtoday{{\sl {Den \number\day. \ifcase\month 
\or Januar\or Februar\or M\"arz\or April\or Mai
\or Juni\or Juli\or August\or September\or Oktober
\or November \or Dezember\fi\space \number\year}}}    
\def\today{{\sl {\ifcase\month
\or January\or February\or March\or April\or May
\or June\or July\or August\or September\or October
\or November \or December\fi \space\number\day,\space 
                                            \number\year}}}

\renewcommand{\a}{\alpha}
\renewcommand{\b}{\beta}
\newcommand{\g}{\gamma}           
\renewcommand{\d}{\delta}         
\newcommand{\e}{\varepsilon}

\newcommand{\m}{\mu}

         \newcommand{\OM}{\Omega}
\newcommand{\p}{\psi}             
           
\renewcommand{\th}{\theta}         \newcommand{\T}{\Theta}
\newcommand{\f}{{\phi}}           
\newcommand{\vf}{{\varphi}}

\renewcommand{\AA}{{\cal A}}

\newcommand{\GG}{{\cal G}}

\newcommand{\SSS}{{\cal S}}

\newcommand{\esp}{\\[3mm]}

\newcommand{\sla}{\raise.15ex\hbox{$/$}\kern -.57em} 
\newcommand{\Sla}{\raise.15ex\hbox{$/$}\kern -.70em}
\def\h{\hbar}

\newcommand{\lp}{\left(}\newcommand{\rp}{\right)}
\newcommand{\lc}{\left[}\newcommand{\rc}{\right]}
\newcommand{\lac}{\left\{}\newcommand{\rac}{\right\}}

\newcommand{\complex}{{\kern .1em {\raise .47ex
\hbox {$\scriptscriptstyle |$}}
    \kern -.4em {\rm C}}}
\newcommand{\real}{{{\rm I} \kern -.19em {\rm R}}}
\newcommand{\rational}{{\kern .1em {\raise .47ex
\hbox{$\scripscriptstyle |$}}
    \kern -.35em {\rm Q}}}
\renewcommand{\natural}{{\vrule height 1.6ex width
.05em depth 0ex \kern -.35em {\rm N}}}

\newcommand{\trace}{{\rm {Tr} \,}}
\newcommand{\half}{\frac{1}{2}}
\newcommand{\pa}{\partial}
\newcommand{\pad}[2]{{\frac{\partial #1}{\partial #2}}}

\newcommand{\dpad}[2]{{\displaystyle{\frac{\partial #1}{\partial #2}}}}
\newcommand{\dfud}[2]{{\displaystyle{\frac{\delta #1}{\delta #2}}}}
\renewcommand{\dfrac}[2]{{\displaystyle{\frac{#1}{#2}}}}
\newcommand{\dsum}[2]{\displaystyle{\sum_{#1}^{#2}}}   
\newcommand{\dint}{\displaystyle{\int}}

\newcommand{\ie}{{{\em i.e.},\ }}

\newcommand{\twiddle}{\lower.9ex\rlap{$\kern -.1em\scriptstyle\sim$}}

\newcommand{\bra}[1]{\left\langle {#1}\right|}
\newcommand{\ket}[1]{\left| {#1}\right\rangle}

\newcommand{\vev}[1]{\left\langle {#1}\right\rangle}
\newcommand{\equ}[1]{(\ref{#1})}
\newcommand{\eq}{\begin{equation}}
\newcommand{\eqn}[1]{\label{#1}\end{equation}}
\newcommand{\eea}{\end{eqnarray}}
\newcommand{\eqa}{\begin{eqnarray}}
\newcommand{\eqan}[1]{\label{#1}\end{eqnarray}}
\newcommand{\ba}{\begin{array}}
\newcommand{\ea}{\end{array}}
\newcommand{\eqac}{\begin{equation}\begin{array}{rcl}}
\newcommand{\eqacn}[1]{\end{array}\label{#1}\end{equation}}

\newcommand{\bz}{\begin{enumerate}}
\newcommand{\ez}{\end{enumerate}}



\begin{document}
\global\long\def\eb{\mathbf{e}}

\global\long\def\A{\boldsymbol{\mathsf{A}}}

\global\long\def\B{\boldsymbol{\mathsf{B}}}


\global\long\def\R{\mathsf{R}}

\global\long\def\T{\mathsf{T}}

\global\long\def\M{\mathcal{M}}

\global\long\def\V{\mathbb{V}}




\global\long\def\eps{\boldsymbol{\varepsilon}}

\global\long\def\bOM{\boldsymbol{\Omega}}

\global\long\def\Vol{\mathrm{Vol}}

\global\long\def\so{\mathfrak{so}(3,1)}

\global\long\def\tr{\mathrm{tr}}

\global\long\def\im{\mathrm{i}}

\global\long\def\pd#1#2{\frac{\partial#2}{\partial#1}}

\global\long\def\td#1#2{\frac{d#2}{d#1}}

\global\long\def\ded#1#2#3{\frac{\delta#3}{\delta A_{\theta}^{#1}#2}}

\global\long\def\inp#1#2{\langle#1\,,#2\rangle}

\global\long\def\is{\!}

\global\long\def\udd#1#2#3{#1^{#2}\is_{#3}}

\global\long\def\dud#1#2#3#4{#1_{#2}\!^{#3}\!_{#4}}

\global\long\def\duu#1#2#3{#1_{#2}\is^{#3}}

\global\long\def\bra#1{\langle#1|}

\global\long\def\ket#1{|#1\rangle}

\title{Loop Quantization of the Supersymmetric \\ Two-Dimensional $BF$ Model}

\author{Clisthenis P. Constantinidis\footnote{Work supported
   in part by the Conselho Nacional de Desenvolvimento Cient\'{\i}fico e
   Tecnol\'{o}gico -- CNPq (Brazil) and 
by the PRONEX project No. 35885149/2006 from FAPES -- CNPq (Brazil).},
Ruan Couto\footnote{Work suported by  CAPES (Brazil)},
Ivan Morales$^{*,}$\footnote{Work supported by the Funda\cao\ de Amparo 
\`a Pesquisa do Esp\ii rito Santo -- FAPES} \\
and Olivier Piguet$^*$\footnote{Present address: Physics Department, 
Federal University of Vi\c cosa -- UFV, Vi\c cosa, MG, Brazil}\\
$\ $\\
{\small Departamento de F\ii sica, Universidade Federal do Espirito Santo (UFES)}\\
{\small Vit\'oria, ES, Brazil}
}

\date{}

\maketitle

\begin{center} 

\vspace{-5mm}

{\small\tt E-mails: cpconstantinidis@pq.cnpq.br, 
ruan.giacomini@gmail.com
 mblivan@gmail.com,
opiguet@yahoo.com}
\end{center}


\begin{abstract}In this paper we consider the quantization of the $2d$ $BF$ model 
coupled to topological matter. Guided by the rigid supersymmetry this system 
can be viewed as a super-$BF$ model, where the field content is expressed in terms 
of superfields. A canonical analysis is done
and the constraints are then implemented at the quantum level in order to construct 
the Hilbert space
of the theory under the perspective of Loop Quantum Gravity methods.

\end{abstract}

\section{Introduction}

It is well known that the $BF$ model in two-dimensional space-time 
for the gauge group SO(1,2)
  is equivalent 
to the Jackiw-Teitelboim model for  two-dimensional 
gravity with cosmological 
constant~\cite{Jackiw:1982hg,Jackiw:1984je,Jackiw:1995hb,Teitelboim:1983ux,Teitelboim:1983fg}, 
and in order to develop methods of quantization for gravity, it has been used as 
a useful laboratory~\cite{Isler:1989hq, Fukuyama:1985gg,
 Livine:2003kn, IvanMTese, Constantinidis:2008br, Constantinidis:2008ty}.  
Leitgeb, Schweda and Zerrouki~\cite{schweda}  have proposed an enlargement  
of the $2d$ $BF$ model, 
in which it is coupled to vector and scalar fields. The action is 
\begin{eqnarray}
\label{topmatt}
S =  \dfrac{1}{2} \dint  d^2x \epsilon^{\mu \nu} \phi^i F_{\mu \nu}^i  
+ \dint d^2x  \epsilon^{\mu \nu} (D_{\mu} B_{\nu})^i\psi^i
\end{eqnarray}
where $\phi^i$ is a scalar field, $F_{\mu \nu}^i$ is the curvature associated 
to the gauge field $A^i$, $\psi^i$ is another scalar field and $B_{\mu}^i$ a 
vector field. $D_\mu$ is the covariant derivative. The index $i=1,2,3$ labels the 
basis of the Lie algebra of the gauge group, taken in an anti-hermitean basis, 
$[T_i, T_j] = \epsilon_{ij}{}^kT_k$. In this paper we consider the gauge group
SU(2), corresponding to Riemannian gravity with a positive cosmological
constant. All  fields are valued in the Lie
algebra su(2). Such a field $\varphi$ is matrix 
$\varphi=\varphi^i T_i$.

Our purpose here is to perform the quantization of this model under the perspective 
of Loop Quantum Gravity. In order to do this we consider the canonical structure of 
the theory, obtain the first class constraints and impose them at the 
quantum level for the construction of the Hilbert space. But we still 
explore another symmetry of the action (\ref{topmatt}), namely, a rigid 
supersymmetry  present in the model, which will guide us through the 
construction of the quantum theory\footnote{This supersymmetry seemed unnoticed 
by the authors of Ref.~\cite{schweda}}.

\section{The supersymmetric $BF$  model}
\label{revisao}
In the present context, $N=1$ supersymmetry transformations are
generated by a unique nilpotent operator $Q$: $Q^2=0$.  
$N=1$ superfields read 
\eq
\vf(x,\th) = \vf_0(x) + \th \vf_1(x)\,,
\eqn{superfield}
where $x=(x^\mu,\,\m=0,1)$ are the spacetime manifold coordinates -- denoted below as
$(t,x)$ -- and $\th$, with $\th^2=0$, is the (unique) Grassman
superspace coordinate. By definition a superfield transforms infinitesimally under
supersymmetry as
\eq
Q\vf = \pad{}{\th}\vf\,,\quad\mbox{or, in components}\,,\quad
Q\vf_0=\vf_1\,,\ Q\vf_1=0\,.
\eqn{susy-transf}
Let us introduce the superfield extensions of the fundamental fields present in 
the usual $2d$ $BF$ model: (whose components are the fields present in the 
Leitgeb-Schweda-Zerrouki action \equ{topmatt}:) 
\eq
\Phi=\psi+\theta \phi\,,\quad
\AA =A+\theta B
\eqn{sa}
where $\Phi$ is an odd parity scalar superfield, and $\cal A$  
the even parity superconnection\footnote{We are considering  
even and odd parities in order 
to distinguish the fields, and other objects, of bosonic and fermionic nature, 
respectively.}. The basic fields transform under supersymmetry as two doublets:
\eq
QA=B\,,\quad QB=0\,,\quad\mbox{and}\quad Q\p=\f\,,\quad Q\f=0\,.
\eqn{susy-transf-fields}
It is easy to check that the Leitgeb-Schweda-Zerrouki action is
invariant under these supersymmetry transformations. This invariance is still
more obvious from the definition \equ{susy-transf}, for the superspace
action
\begin{eqnarray}
\label{St1}
\mathcal{S}_T[\Phi,\cal A]&:=&\mbox{Tr} \dint d\theta \Phi \mathfrak{F}[\cal A], 
\end{eqnarray}
which is equivalent to \equ{topmatt},  
with the difference that here we are considering the fields $B$ and $\psi$ 
with  odd parity. $\mathfrak{F}$ is the supercurvature of the
superconnection $\cal A$:
\begin{eqnarray}
\mathfrak{F}&=&d {\cal A} + \dfrac{1}{2} [{\cal A},{\cal A}] 
=F+\theta\mathbb{F},
\end{eqnarray}
with  $F$ being the usual Yang-Mills curvature and $\mathbb{F}$ its supersymmetry 
partner, and thus, an odd quantity: 
\begin{equation}
F=dA+\dfrac{1}{2}[A,A] , \; \;   \mathbb{F}=DB,
\end{equation}
where $D$ is the covariant derivative\footnote{The brackets $[\cdot,\cdot]$ are
graduated commutators, \ie an anticommutator if both its arguments are odd, 
and a commutator otherwise.}, 
$D=d+[A,\;]$, and $d=dx^\m\pa_\m$ the usual spacetime exterior derivative.  
The trace symbol $\trace$ is taken to be the Killing form of the $su(2)$ algebra. 
Integration in $\th$ is defined by the Berezin integral~\cite{berezin, corn}, which in the 
present case amounts to the definition:
\[
\dint d\th\cdots = \pad{}{\th}\cdots\,,\quad\mbox{or:}\quad \dint d\th\, 1=0\,,\quad
\dint d\th\, \th = 1\,.
\]
Varying the action  (\ref{St1}) we have 
\eq\ba{l}
\delta \mathcal{S}_T=\mbox{Tr} \dint d \theta (\delta\Phi \mathfrak{F}
+\Phi \delta\mathfrak{F})
=\mbox{Tr} \dint d \theta (\delta\Phi \mathfrak{F}+\delta {\cal A} 
\mathcal{D}\Phi)\,,\esp
\phantom{\delta \mathcal{S}_T}
=\mbox{Tr} \dint (\delta\phi F - \delta\psi \mathbb{F} -\delta A(-d\phi-[A,\phi]+[B,\psi])
+ \delta B(D\psi+[A,\psi]))\,,
\ea\eqn{var1}
 where $\mathcal{D}:=d+[\mathcal{A}, \quad ]$ is the covariant derivative for the 
superconnection.  From \equ{var1}
 we obtain the equations of motion
\eq\ba{ll}
\dfrac{\delta S_T}{\delta\phi}=F=0\,,\quad 
&\dfrac{\delta S_T}{\delta A}=d\phi+[A,\phi]-[B,\psi]=D\phi-[B,\psi]=0\,,\esp
\dfrac{\delta S_T}{\delta\psi}=\mathbb{F}=0\,,\quad 
&\dfrac{\delta S_T}{\delta B}=d\psi+[A,\psi]=D\psi=0\,.
\ea\eqn{emt1}
    
\subsection{The gauge group}\label{gauge group}
In order to treat the graded structure of the superfields, we also consider the 
``supergauge group'' $\mathbf{G}$  of elements:
\eq
\GG(x)=\GG(\alpha,\beta):=e^{\Omega(x)}=e^{\alpha(x)+\theta\beta(x)}\,,
\eqn{sgt}
where the parameters $\alpha$ and $\beta$ are evaluated in  $su(2)$. 
Observe that $\alpha$ is even while the quantity $\beta$  is odd, 
 the latter being the supersymmetry transform \equ{susy-transf}
of the former:
$\b=Q\a$. 
When expanded in $\th$, this expression can be written as
\begin{equation}
\label{sgt1}
\GG (\alpha,\beta):=g(\alpha)+\theta\beta\rhd g(\alpha)
\end{equation}
with
\eq
g(\alpha):=e^\alpha=\sum_{n=0}^\infty\dfrac{\alpha^n}{n!} 
\eqn{g(alpha)}
being an element of  $SU(2)$,  and the quantity 
\begin{eqnarray}
\label{ig}
\beta\rhd g(\alpha):=
\b^i\dpad{}{\a^i} g(\a) =
\sum_{n=1}^\infty\dfrac{1}{n!}\sum_{k=1}^n\alpha^{n-k}\beta\alpha^{k-1}\,,
\end{eqnarray}
is defined as the insertion of the Grassmannian parameter 
$\beta$ into  $g$. 
 It is the supersymmetry transform of the group element $g(\a)$,
\eq
\beta\rhd g(\alpha) = Qg(\a) \,,
\eqn{susy-var-of-g}
There exists an inverse element $\GG ^{-1}$, given by
\begin{eqnarray}
\label{sgt2}
G^{-1}(\alpha,\beta):=e^{-\Omega}=e^{-\alpha-\theta\beta}=g^{-1}(\alpha)+\theta\beta\rhd g^{-1}(\alpha)
\end{eqnarray}
with 
\begin{eqnarray}
\beta\rhd g^{-1}(\alpha):=\sum_{n=1}^\infty\dfrac{(-1)^n}{n!}\sum_{k=1}^n\alpha^{n-k}\beta\alpha^{k-1}
\end{eqnarray}
This  insertion enjoys the following property,
\begin{eqnarray}
(\beta\rhd g)g^{-1}=-g(\beta\rhd g^{-1})\,.
 \end{eqnarray}
 Under left and right multiplication of a group element $g$,
\[
g' = g_1 g g_2^{-1}\,,\quad\mbox{with}\ g=g(\a)\,,\ g_k=g(\a_k)\,,\
\b_k=Q\a_k\,,\ k=1,2\,,
\] 
the $\b$-insertion transforms as
\eq
(\b\rhd g)' = g_1(\b\rhd g)g_2^{-1} + (\b_1\rhd g_1)g g_2^{-1}
- g_1 g g_2^{-1} (\b_2\rhd g_2) g_2^{-1} \,.
\eqn{insertion-transf}

\subsection{Symmetries}\label{symmetries}
   As in the usual $BF$ theory, one verifies that the action (\ref{St1}) 
is invariant under the following finite supergauge transformations:   
 for the superconnection $\AA$,
\eq
\label{tcr1}
{\cal A}(x) \rightarrow {\cal A}'(x)=\GG (x)d\GG ^{-1}(x)
+\GG (x){\cal A} (x)\GG (x)^{-1}\,,
\end{equation}
from which we identify
\begin{eqnarray}
\label{tcr2}
A'(x)&=&g(x)dg^{-1}(x)+g(x)A(x)g^{-1}(x)\\
\label{tcr3}
B'(x)&=&g(B+D(g^{-1}(\beta\rhd g))g^{-1}.
\end{eqnarray}
where $D$   is the covariant derivative: 
\begin{eqnarray}
D(g^{-1}(\beta\rhd g))=d(g^{-1}(\beta\rhd g))+Ag^{-1}(\beta\rhd g)+g^{-1}(\beta\rhd g)A\,;
\end{eqnarray}
and   for the supersymmetric extension  $\Phi$ of the scalar field, 
\begin{equation}
\label{tcr4}
\Phi (x) \rightarrow \Phi'(x)=\GG (x)\Phi(x)\GG ^{-1}(x)\,,
\end{equation}
from which one reads
\begin{eqnarray}
\label{tcr5}
\phi'(x)&=&g(x)\phi(x)g^{-1}(x)\\
\label{tcr6}
\psi'(x)&=&g(x)\lp \psi(x)+\theta\lc\psi(x),g^{-1}(x)(\beta\rhd g)(x)\rc\rp g^{-1}(x)
\end{eqnarray}
In order to obtain the infinitesimal supergauge transformations, we take $\Omega(x)$ small in 
 (\ref{sgt}), and from   (\ref{tcr1}) and (\ref{tcr4}), we get
 \begin{eqnarray}
 \label{tgt1}
 \delta_\Omega {\cal A}  &=& - \mathcal{D} \Omega ,\\
\label{tgt2}
 \delta_\Omega \Phi&=&-[\Phi,\Omega].
 \end{eqnarray}
Considering the splitting  in even and odd quantities yields 
the following infinitesimal gauge transformations, which are of two kinds:
\begin{itemize}
\item  gauge transformations of type $\alpha$, which involve the even parameter:
\begin{eqnarray}
\label{tgt5}
\delta_\alpha A=-D\alpha\,,\quad
\delta_\alpha \phi=[\alpha,\phi]\,,\quad
\delta_\alpha B=[\alpha,B]\,,\quad
\delta_\alpha \psi=[\alpha,\psi]\,,
\end{eqnarray}
\item gauge transformations of type $\beta$, associated to the odd parameter:
\begin{eqnarray}
\delta_\beta A=0\,,\quad
\delta_\beta \phi=[\beta,\psi]\,,\quad
\delta_\beta B=D\beta\,,\quad
\delta_\beta \psi=0\,.
\end{eqnarray}
\end{itemize}
These  gauge transformations 
are the same as the ones of the Leitgeb-Schweda-Zerrouki model, given by 
the action (\ref{topmatt}).  
The difference is that the transformations of parameter $\b$ are now interpreted as local 
supersymmetry transformations, $\b$ being now odd. 

Diffeomorphisms are also symmetries of the $BF$ model. 
Considering   a vector field $v$, infinitesimal diffeomorphism transformations,
$\d x^\m=v^\m(x)$, are given by the Lie derivative
\begin{eqnarray}
\mathcal{L}_v\Phi=i_v d\Phi\,,\quad
\mathcal{L}_v{\cal A}=(i_vd+d i_v){\cal A},
\end{eqnarray}
  where $i_v$ is the interior derivative in the direction of the vector $v$. 
We can easily check that these infinitesimal diffeomorphisms can be expressed as  
 \begin{eqnarray}
 \mathcal{L}_v\Phi
=i_v\dfrac{\delta S_T}{\delta B}+\theta i_v\dfrac{\delta S_T}{\delta A}
-\delta_{(i_v{\cal A})}\Phi\,,\quad
 \mathcal{L}_v {\cal A}
=i_v\dfrac{\delta S_T}{\delta \phi}+\theta i_v\dfrac{\delta S_T}{\psi}
-\delta_{(i_v{\cal A})}{\cal A}\,,
 \end{eqnarray}
from which one realizes that they are related to gauge symmetries --   
if we consider $i_v{\cal A} = v^\mu {\cal A}_\mu$ as the parameter 
of the infinitesimal transformation --
 modulo equations of motion.

\section{Canonical analysis}
In order to proceed with the canonical analysis, we consider the 
two-dimensional spacetime  manifold foliated as  $\mathcal{M}=\mathbb{R}\times \Sigma$,
with coordinates  $x^\mu=(t,x)$, where $t$ 
and  $x$ are time and space coordinates respectively.  
The action (\ref{St1}) can be written as 
\begin{eqnarray}
S_T&=&\dint_\mathbb{R} dt \dint_\Sigma dx d\theta \Phi_i(\partial_t {\cal A}_x^i-\partial_x {\cal A}_t^i+f_{jk}\,\!^i{\cal A}_t^j{\cal A}_x^k) \nonumber\\
&=&\dint_\mathbb{R} dt \dint_\Sigma dx \,\mathcal{L}_T,
\label{action-2}\end{eqnarray}
with 
\begin{eqnarray}
\mathcal{L}_T= \phi_i(\partial_t A_x^i-\partial_x A_t^i+f_{jk}\,\!^iA_t^jA_x^k)-\psi_i(\partial_t B_x^i-\partial_x B_t^i+f_{jk}\,\!^i(A_t^jB_x^k-A_x^jB_t^k))
\label{lagdensity}
\end{eqnarray}
where in the second equality we have integrated over   the Grassmannian parameter $\theta$, using the Berezin integral.

Before proceeding with the analysis, we define the generalization of the 
Poisson brackets for the case where we have fields with even and odd 
parities~\cite{henn}, 
so we define, for any quantities $M$ and $N$\footnote{The parity function, denoted 
by $|\ \ |$ is defined as: $|M|=0\,(1)$ if $M$ 
is  even (odd).},
\begin{eqnarray}
\label{7}
\left\{M,N\right\}:=\dsum{A}{}
\dint dx (-1)^{\vert M\vert\vert \mathcal{Q}_ A\vert}\left(\dfrac{\delta M}
{\delta \mathcal{Q}_ A(x)}\dfrac{\delta N}{\delta\mathcal{P}^ A(x)}
-(-1)^{\vert \mathcal{Q}_ A\vert}\dfrac{\delta M}
{\delta\mathcal{P}^ A(x)}\dfrac{\delta N}{\delta \mathcal{Q}_ A(x)}\right),
\end{eqnarray}
where $\mathcal{Q}_ A$ and $\mathcal{P}^ A$ are generalized configuration 
variables and their conjugate momenta. In particular, the basic
non-vanishing brackets are
\[
\lac \mathcal{Q}_ A(x),\,\mathcal{P}^B(y) \rac 
= (-1)^{|\mathcal{Q}_ A||\mathcal{P}^B|} \d^B_A \d(x-y)\,.
\]
Taking $A^i_x$, $A^i_t$, $B^i_x$ and $B^i_t$ as configuration variables,
we read from the action (\ref{action-2},\ref{lagdensity}) that $\f_i$
and $\p_i$ can be identified with the conjugate momenta of $A^i_x$ and
$B^i_x$, respectively. 

Thus $A_x$ and $B_x$ have non-vanishing Poisson brackets with $\f$ and
$\p$, respectively:
\eq\ba{lll}
\left\{A_x^i(x),\phi_j(y)\right\}&=\delta^i_j\delta(x-y)
&=-\left\{\phi_j(y),A_x^i(x)\right\}\,,\esp
\left\{B_x^i(x),\psi_j(y)\right\}&=-\delta^i_j\delta(x-y)
&=\left\{\psi_j(y),B_x^i(x)\right\}\,.
\ea\eqn{scpt1}
On the other hand, the conjugate momenta 
$\pi_i^{(A_t)}$ and $\pi_i^{(B_t)}$ of
$A^i_t$ and $B^i_t$ turn out to vanish: in the Dirac-Bergmann canonical
formalism~\cite{dirac}, these are interpreted as primary constraints:
\eq
\pi_i^{(A_t)}(x) \approx 0\,,\quad \pi_i^{(B_t)}(x) \approx 0\,,
\eqn{prim-constr}
where $\approx$ means a weak equality, \ie an equality which will be
fulfilled only after of all Poisson algebra manipulations are done.
For consistency under the time evolution generated by the canonical 
Hamiltonian, these constraints produce secondary constraints.
The canonical Hamiltonian, obtained from the Lagrangian  \equ{lagdensity} by a
Legendre transform, reads
\begin{eqnarray}
\label{ht3}
H=-\dint dx \lp A_t^i(x)\mathbb{G}_i(x)+B_t^i(x)\mathbb{S}_i(x)\rp.
\end{eqnarray}
with
\eq
\mathbb{G}_i(x) :=D_x\phi_i(x)+[B_x(x),\psi(x)]_i \,,\quad
\mathbb{S}_i(x) :=D_x\psi_i(x) \,,
\eqn{vt1}  
and the secondary constraints are the terms proportional to $A_t$ and
$B_t$ in the Hamiltonian (which is thus itself a constraint, as expected 
in a background independent theory):
\eq
\mathbb{G}_i(x) \approx 0\,,\quad \mathbb{S}_i(x) \approx 0\,.
\eqn{sec-constraints}
These constraints are first class, \ie they have weakly vanishing Poisson
brackets.
 Indeed, let us use the smeared form of these constraints for more
clarity:
\begin{eqnarray}
\mathbb{G}(\alpha):=\dint dx\, \alpha^i(x)\mathbb{G}_i(x)\,,\quad
\mathbb{S}(\beta):=\dint dx\, \beta^i(x)\mathbb{S}_i(x)\,,
\end{eqnarray}
where $\a^i$ and $\b^i$ are smooth test functions. $\mathbb{S}$ and $\mathbb{G}$
form a supersymmetry doublet: $Q\mathbb{S}$ $=$ $\mathbb{G}$. Putting $\mathbb{S}(x)$ and 
$\mathbb{G}(x)$ 
together  into a superfield $\SSS(x,\theta) = \mathbb{S}(x) + \theta \mathbb{G}(x)$, 
and writing, with $\OM$ as given in \equ{sgt},  
\eq
\SSS(\OM) = \dint dx\dint d\theta\, \OM^i(x,\theta) \SSS_i(x,\theta)
= \mathbb{G}(\a) + \mathbb{S}(\b) \approx 0\,,
\eqn{constraints}
one checks that the latter form a closed Poisson algebra:
\[
\lac \SSS(\OM_1),\,\SSS(\OM_2\rac = \SSS([\OM_1,\OM_2])\,.
\]
Being thus first class, $\SSS$ is the infinitesimal  generator of 
gauge transformations, namely of the supertransformations
(\ref{tcr1},\ref{tcr4}):
\eq\ba{lll}
\lac \SSS(\OM),\,\AA_x(x,\theta)\rac &=& \pa_x\OM(x,\theta) +
[\AA_x(x,\theta),\,\OM(x,\theta)]\,,\esp
\lac \SSS(\OM),\,\Phi(x,\theta)\rac &=& 
[\Phi(x,\theta),\,\OM(x,\theta)]\,,\esp \,.
\ea\eqn{supertransf}
In components, this yields the algebra
\begin{eqnarray}
\{\mathbb{G}(\alpha),\mathbb{G}(\alpha')\}=\mathbb{G}([\alpha',\alpha])\,,\quad
\{\mathbb{G}(\alpha),\mathbb{S}(\beta)\}=\mathbb{S}([\alpha,\beta])\,,\quad
\{\mathbb{S}(\beta),\mathbb{S}(\beta')\}=0\,,
\end{eqnarray}
and the infinitesimal form of the gauge transformations of 
type $\alpha$ and type $\beta$ 
given in Subsection \ref{symmetries}:
\begin{eqnarray}
\label{tgt13}
&\{\mathbb{G}(\alpha),A^i_x(x)\}=D_x\alpha^i(x)\,,\quad
\{\mathbb{G}(\alpha),\phi_i(x)\}=-[\alpha(x),\phi(x)]_i\,,\nonumber\\
&\{\mathbb{G}(\alpha),B^i_x(x)\}=-[\alpha(x),B_x(x)]^i\,,\quad
\{\mathbb{G}(\alpha),\psi_i(x)\}=-[\alpha(x),\psi(x)]_i\,,
\end{eqnarray}
\begin{eqnarray}
&\{\mathbb{S}(\beta),A^i_x(x)\}=0\,,\quad
\{\mathbb{S}(\beta),\phi_i(x)\}=-[\beta(x),\psi(x)]_i\,,\nonumber\\
&\{\mathbb{S}(\beta),B^i_x(x)\}=D_x\beta^i\,,\quad
\{\mathbb{S}(\beta),\psi_i(x)\}=0\,.
\end{eqnarray}


\subsection*{Classical observables}
Once we know all the transformation rules for  the holonomies and the insertions, 
we search for the gauge invariant quantities of the theory, the observables.
Considering the topology of $\Sigma$ being $S_1$, it is well known that for the 
two-dimensional $BF$ model one observable (gauge invariant quantity) is the Wilson loop, 
$W_0$,~\cite{Livine:2003kn, Constantinidis:2008ty}, the trace of 
the holonomy trough a closed path  -- which here coincides with space $S_1$ itself.
 In this model we construct another 
observable, $W_1$, which is the trace of the insertion of $B$ in the 
holonomy along the same closed path. Observe that $W_0$ is an even quantity and $W_1$ is odd. 
We write them as 
\eq
W_0:=\mbox{Tr}(h[A])\,,\quad
W_1:=\mbox{Tr}(B\rhd h[A])\,.
\eqn{ott1}
Their gauge invariance follows from (\ref{ht1}) e (\ref{sht7}).
We note that there are no gauge invariant multilinear $B$-insertion of
order $\geq2$ in $B$.
Other invariant quantities are
\eq
L_0:=\mbox{Tr}(\psi\phi)=\psi^i\phi_i\,,\quad
L_1:=\mbox{Tr}(\f^2)=\f^i\f_i\,.
\eqn{ott3}
 $L_1$ is present in the   two-dimensional $BF$ 
model~\cite{Livine:2003kn,Constantinidis:2008ty}. One observes that
these four observables form two supersymmetry doublets:
\[
QW_0=W_1\,,\quad QW_1=0\,,\quad\quad QL_0=L_1\,,\quad QL_1=0\,.
\]
The first two can be taken as the basis for the wave functionals which 
will describe the quantum space of the model.

\section{Loop quantization}
We now begin with the construction of the Hilbert space  of the theory, 
in which the basic elements are wave functionals of the type  
$\Psi[\AA]:=\Psi[A,B]$.  We thus choose as coordinates  
the connection $A$ and the vector field $B$, which fix the polarization as
\begin{eqnarray*}
\hat{A_x}(x)\Psi[A,B]:=A_x(x)\Psi[A,B],\quad \hat{\phi}(x)\Psi[A,B]:=i\hbar\dfrac{\delta}{\delta A_x}\Psi[A,B],\\
\hat{B_x}(x)\Psi[A,B]:=B_x(x)\Psi[A,B],\quad \hat{\psi}(x)\Psi[A,B]:=i\hbar\dfrac{\delta}{\delta B_x}\Psi[A,B],
\end{eqnarray*}
and the brackets (\ref{scpt1})  are promoted to  (graded) commutators; the non-vaishing ones are
\begin{eqnarray}
\left[\hat{A}_x^i(x),\hat{\phi}_j(y)\right]_-=i\hbar\delta^i_j\delta(x-y)\,,\quad
\left[\hat{B}_x^i(x),\hat{\psi}_j(y)\right]_+=-i\hbar\delta^i_j\delta(x-y)
\end{eqnarray}

\subsection{Superholonomies}

 Following the steps of Loop Quantum Gravity, in order to achieve a well
defined Hilbert space, we shall write the wave functional 
in terms of the holonomies of the connection and of their 
$B$-insertions -- to be defined hereafter --
instead of the local fields $A$ and $B$ themselves. 
The constraints are then imposed  in order to select the vectors belonging to
the physical Hilbert  space
of the theory.  Holonomies are convenient variables when one 
imposes the Gauss constraint, once they are endowed with transformation 
properties  which permit to construct gauge invariant quantities in a 
relatively simple way.   

Guided by the (rigid) supersymmetry of the model, 
we first construct the superholonomy for the superconnection $\AA$, trough a path 
$\gamma$ on the manifold $\Sigma$, as follows:
\begin{eqnarray}
\label{sht1}
\mathbf{H}_\gamma[\AA]:=\mathbf{H}_\gamma[A,B]:=\mathcal{P}e^{-\int_\gamma \AA}
=\mathcal{P} e^{-\int_\gamma(A+\theta B)}.
\end{eqnarray}
Expanding in powers of $\th$, and rearranging terms this expression can be written as 
\begin{eqnarray}
\label{sht2}
\mathbf{H}_\gamma[A,B]:=h_\gamma [A]-\theta B\rhd h_\gamma [A]
\end{eqnarray}
where $h_\gamma[A]$ is the usual holonomy for the connection $A$ through  $\gamma$ on $\Sigma$:
\eq
h_\gamma[A]=Pe^{-\int_\gamma A}\,,
\eqn{hta1}
and $B\rhd h_\gamma [A]$ 
is the insertion of the Grassmannian field  $B$  in the holonomy  through the curve
$\gamma$ parametrized by $s$,  with  $s_0<s<s_f$, which reads
\eq
B\rhd h_\gamma [A](s_f,s_0):=\dint_{s_0}^{s_f} ds\, h[A](s_f,s)B_x(s)h[A](s,s_0)\,.
\eqn{iht1}
Making use of the notion of superfields, the  superholonomy $\mathbf{H}_\gamma[A,B]$ 
can be written as 
\begin{eqnarray}
\label{sht3}
\mathbf{H}_\gamma[A,B]:=h_\gamma [A] + \theta Q h_\gamma [A],
\end{eqnarray}
where   $Q$ is the supersymmetry generator \equ{susy-transf},
and we consequently have
\begin{equation}
\label{sht4}
Q h_\gamma[A]= -B\rhd h_\gamma [A]\,.
\end{equation}

With these considerations in mind, let us now introduce some useful relations present in this formalism. Consider a composed path on  $\Sigma$, given by $\gamma=\gamma_2\circ\gamma_1$. 
The holonomy  $h_\gamma[A](s_f,s_0)$ satisfies the following property, 
\begin{equation}
h_\gamma[A](s_f,s_0)=h_{\gamma_{2}}[A](s_f,s)\, h_{\gamma_1}[A](s,s_0).
\label{holocomp}
\end{equation}
where $s_0$ and $s_f$ are the initial and final points of the path $\g$ and $s$ is the end point of $\g_1$ and initial point of $\g_2$.
The insertion of the field $B$ in the holonomy, \ie  the quantity $Qh_\gamma[A](s_f,s_0)$, satisfies the following one
\begin{equation}
\label{sht5}
Q(h_\gamma[A](s_f,s_0))=Qh_{\gamma_2}[A](s_f,s)\, h_{\gamma_1}[A](s,s_0)
+h_{\gamma_2}[A](s_f,s)\, Qh_{\gamma_1}[A](s,s_0).
\end{equation}
\label{sht5-1}
Under gauge transformations holonomies transform as follows:
\begin{equation}
\label{ht1}
h'_\gamma[A](s_f,s_0)=g(s_f)h_\gamma[A](s_f,s_0)g^{-1}(s_0),
\end{equation}
and in a similar fashion the supergauge transformation for the superconnection is given by
\begin{eqnarray}
\label{sht6}
\mathbf{H}'_\gamma[\AA](s_f,s_0):=
\GG (s_f)\mathbf{H}_\gamma[\AA](s_f,s_0)\GG ^{-1}(s_0),
\end{eqnarray}
where $\GG $ is parametrized as in  (\ref{sgt1}). 
 Differentiating this expression in $\th$ and using  
(\ref{sgt2}), (\ref{sht4}), (\ref{sht3}) and (\ref{ht1}),
we obtain the form of the gauge transformations for the $B$-insertion,
\eq\ba{l}
(B\rhd h_\gamma [A](s_f,s_0))'
=g(s_f) B\rhd h_\gamma [A](s_f,s_0) g^{-1}(s_0)\esp
 \quad\quad\quad+ (\beta\rhd g)(s_f)h_\gamma[A](s_f,s_0) g^{-1}(s_0)
 - g(s_f) h_\gamma[A](s_f,s_0)
(g^{-1}\beta\rhd g) g^{-1}(s_0)\,.
\ea\eqn{sht7}
{\bf N.B.} This transformation rule has the same form as that of the
$\b$-insertion given in \equ{insertion-transf}, but with $g_1$, $g_2$
replaced by $g(s_f)$, $g(s_0)$, $g$ with $h_\g[A]$ and $\b\rhd g$
with the $B$-insertion $B\rhd h_\g[A]$.

\subsection{Hilbert space}\label{hilbert-space}

 The Hilbert space of the theory will be constructed on the basis of 
wave  functionals of the form
\eq
\Psi[A,B]:=f(h[A],B\rhd h[A])\,,
\eqn{cyl-vector}
which are the so-called {\sl  cylindrical functions}. Their arguments are  
the holonomies $h[A]$ and the $B$-insertions $B\rhd h [A]$, 
given in  (\ref{hta1}) and (\ref{iht1}) respectively, with $\g$ 
a curve in the space $S_1$.  The set of
such functionals forms the vector space Cyl.

The physical state vectors are then obtained by the imposition of  
the constraints (\ref{vt1}). Thus, the physical wave functionals 
will be gauge invariant, hence given by functions of the gauge invariant
quantities $W_0$ and $W_1$ given by \equ{ott1}:
\eq
\Psi[A,B] = f(W_0[A],W_1[A,B]) = \psi(h[A],B\rhd h[A])\,.
\eqn{wave-funct}
The second equality defines the wave functional as a function $\p$
 of the hol\-on\-omies and their $B$-insertions, \ie a function on the
supergauge group ${\bf G}$.

Due to $W_1$ being an anticommuting odd number, $W_1^2=0$, the function
$f$ in \equ{wave-funct}
expands in $W_1$ as  
\eq
f(W_0[A],W_1[A,B]) = a(W_0) + W_1 b(W_0)\,.
\eqn{expansion-f}
Choosing a primitive $\hat b$ of the function $b$, $b={\hat b}'$, 
we can rewrite the last equation as 
\eq
f(W_0[A],W_1[A,B]) = a(W_0[A]) + W_1[A,B] {\hat b}'(W_0[A])
= a(W_0[A]) + Q\hat b(W_0[A])\,,
\eqn{expansion-f'}
which shows that the space of supergauge invariant functionals splits
into 
singlet and doublet representations of the rigid supersymmetry.
The singlets are the constant functions.

We also conclude from \equ{expansion-f} that we have two types of wave functionals:
\eq\ba{ll}
\mbox{even:}\quad &\Psi_+[A,B] = f_+(W_0[A]) = \psi_+(h[A])\,,\esp
\mbox{odd:}\quad &\Psi_-[A,B] = Qf_-(W_0[A]) = 
\trace(B\rhd h[A])\,\psi_-(h[A]) = W_1 \psi_-(h[A]) \,.
\ea\eqn{Psi+-}
The internal product of two state vectors \equ{wave-funct} will be
defined by an integral on the supergauge group ${\bf G}$ of elements
$\GG$:
\eq
\vev{\Psi_1|\Psi_2} = \dint_{\mathbf{G}} d\m(\GG) 
\lp\psi_1(\GG)\rp^* \psi_2(\GG)\,,
\eqn{int-product-group}
where $d\m$ is an invariant measure on ${\bf G}$, a generalization of the
usual Haar measure we are going to give now.

\subsubsection{Integration measure}\label{integration measure}

In order to define an internal product, we need to define an integration
measure on the configuration space, whose points are holonomies of $A$ and
$B$-insertions. For this purpose we will now construct an invariant integration measure on 
the supergauge group 
${\bf G}$ defined in Subsection \ref{gauge group}:
\eq
\dint_{\bf G} d\mu(\GG)f (\GG)
=\dint_{\bf G} d^3\alpha d^3\beta\,\rho(\alpha,\beta)F(\alpha,\beta)\,,
\eqn{imt1}
where $d^3\b = \pa_{\b^3}\pa_{\b^2}\pa_{\b^1}$ is the Berezin
integration measure over the odd parameters of the group, 
$F(\alpha,\beta)=f(\GG(\a,\b))$ and $\rho(\alpha,\beta)$ is a weight
which must be chosen such that the integral be invariant under left and
right multiplication:
\eq
\dint_{\bf G} d\mu(\GG)f (\GG) =
\dint_{\bf G} d\mu(\GG)f (\GG_1\GG) = \dint_{\bf G} d\mu(\GG)f (\GG\GG_1) \,,
\quad \forall \GG_1 \in {\bf G}\,.
\eqn{invar-integral}
 This will ensure the supergauge invariance of the
scalar product.

In terms of the parametrization $(\a,\b)$ defined by \equ{sgt}, the
product $\GG'=\GG_1\GG$ writes
\eq
\a'{}^i=p^i(\a_1,\a)\,,\quad \b'{}^i= Q\,p^i(\a_1,\a) = 
\lp \b^j_1\pa_{\a^j_1} + \b^j\pa_{\a^j} \rp p^j(\a_1,\a)\,,
\eqn{product-law}
where $p(\a_1,\a)$ is the product law, in terms of the $\a$
parametrization,
for the bosonic part of the supergauge group, here SU(2). 
Thus, the invariance condition (we restrict ourselves here to the left
invariance)
\eq
\dint_{\bf G} d^3\alpha d^3\beta\,\rho(\alpha,\beta)F(\alpha,\beta)
= \dint_{\bf G} d^3\alpha d^3\beta\,\rho(\alpha,\beta)F(\alpha',\beta')\,,
\eqn{inv-cond-parameters}
implies, thanks to the super-Jacobian of the change of integration variables
$(\a,\b)\to(\a',\b')$ in the right-hand side being equal to one, the
invariance condition
\eq
\rho(\a,\b) = \rho(\a',\b')
\eqn{inv-rho}
for the integration weight.

The $\b$-dependence of the function $F(\a,\b)$ may have
any one of the four following forms:
\eq
F_0(\a)\,,\quad
\mbox{or}\quad F_{i}(\a)\b^i\,,\quad \mbox{or}\quad \frac12 F_{ij}(\a)\b^i\b^j  
\,,\quad \mbox{or}\quad \frac{1}{3!} F_{ijk}(\a)\b^i\b^j\b^k\,.
\eqn{F-exp}
Each of these expressions must be integrated with one of the following weight
functions $\rho(\a,\b)$, respectively:
\eq
 \frac{1}{3!}\rho_{ijk}(\a)\b^i\b^j\b^k\,,\quad \mbox{or}\quad  
 \frac12\rho_{ij}(\a)\b^i\b^j\,,\quad \mbox{or}\quad  
 \rho_{i}(\a)\b^i \,,\quad \mbox{or}\quad 
\rho_0(\a)\,.
\eqn{rho-exp}
The integrals read, after the Berezin integration 
$\int d^3\b \b^i\b^j\b^k = \e^{ijk}$ has been performed:
\eq\ba{l}
\dfrac{1}{3!}\e^{ijk}\dint d^3\a\,\rho_{ijk}(\a) F_0(\a) 
=  \dint d^3\a\,\tilde\rho(\a) F_0(\a)  \,, \esp
\dfrac{1}{2}\e^{ijk}\dint d^3\a\,\rho_{ij}(\a) F_{k}(\a) 
=  \dint d^3\a\,{\tilde\rho}^i(\a) F_i(\a)   \,,\esp
\dfrac{1}{2}\e^{ijk}\dint d^3\a\,\rho_{i}(\a) F_{jk}(\a) 
=  \dfrac{1}{2} \dint d^3\a\,{\tilde\rho}^{ij}(\a) F_{ij}(\a) \,, \esp
\dfrac{1}{3!}\e^{ijk}\dint d^3\a\,\rho_0(\a) F_{ijk}
= \dfrac{1}{3!} \dint d^3\a\,{\tilde\rho}^{ijk}(\a) F_{ijk}(\a)\,.
\ea\eqn{integrals}
Writing the invariance condition \equ{inv-rho} for each of the weight
functions \equ{rho-exp}, one obtains terms in $\b$ and $\b_1$. Those in
$\b_1$ are irrelevant 
since they are of lower order in $\b$ and thus do not contribute to the
Berezin integrals. This finally yields the conditions
\eq\ba{lll}
{\tilde\rho}(\a') &=& {\tilde\rho}(\a)\, {\rm det}\dpad{\a}{\a'} 
\,,\esp
{\tilde\rho}^i(\a') &=& {\tilde\rho}^m(\a)\, \dpad{{\a'}^i}{\a^m}\,{\rm det}\dpad{\a}{\a'}
\,,\esp
{\tilde\rho}^{ij}(\a') 
&=& {\tilde\rho}^{mn}(\a)\, \dpad{{\a'}^i}{\a^m}\dpad{{\a'}^j}{\a^n}
\,{\rm det}\dpad{\a}{\a'} \,,\esp
{\tilde\rho}^{ijk}(\a') 
&=& {\tilde\rho}^{mnp}(\a)\, \dpad{{\a'}^i}{\a^m}\dpad{{\a'}^j}{\a^n}\dpad{{\a'}^k}{\a^p}
\,{\rm det}\dpad{\a}{\a'}\,.
\ea\eqn{invariances-cond}
Since the wave functionals we consider
are linear in the $B$-insertions, it suffices, for defining an internal product, to
restrict the remainder of the discussion to the terms of at most order 2 in
$\b$. Thus we are interested to find solutions for weights $\tilde\rho$ 
corresponding to the first three lines of \equ{integrals} and of \equ{invariances-cond}.
The first of conditions \equ{invariances-cond} 
means that ${\tilde\rho}$ transforms like the Haar
measure of the bosonic part of the supergauge group, and can thus be
identified with it:
\eq
{\tilde\rho}(\a) = \rho_{\rm H}(\a)\,,
\eqn{rho_3}
The second condition implies that $\pa_{\a^i}{\tilde\rho}^i$ transforms
like the Haar measure $\tilde\rho$, hence ${\tilde\rho}^i$ can be
identified with a solution of
\eq
\pa_{\a^i}\tilde\rho^i(\a) = \rho_{\rm H}(\a)\,.
\eqn{div-tilde-rho}
The third condition implies that the divergence $\pa_{\a^i}{\tilde\rho}^{ij}$ transforms
like  $\tilde\rho{}^j$, hence we  are tempted to identify it with the latter. 
However this does not work because this would imply the vanishing of
$\rho_{\rm H}$ as can be seen taking the divergence of both terms and 
observing that $\pa_{\a^j}\pa_{\a^i}{\tilde\rho}^{ij}$ identically
vanishes due to the antisymmetry of ${\tilde\rho}^{ij}$. We need
another solution  for the latter, obeying the third of
the invariance conditions \equ{invariances-cond}. But we do not need to
know it explicitly. Let us suppose that we have such a solution for ${\tilde\rho}^{ij}$. 
As we shall see in Subsection \ref{internal-product}, Eq. \equ{odd-wave},
supergauge invariant wave functions are either functions of $\a$, or
linear in $\b$, of the form $\b^i\pa_i f(\a)$. Thus
we only need to define 
integrals of the restricted form
\eq
\dint d^3\a \tilde\rho{}^{ij}(\a)\pa_{\a^i}f_1(\a) \pa_{\a^j}f_2(\a)\,,
\eqn{prod-two-b-insert} 
which, after partial integrations, is equal to
\[
- \dint d^3\a \pa_{\a^j}\tilde\rho{}^{ij}(\a)\pa_{\a^i}f_1(\a) f_2(\a)
= \dint d^3\a \pa_{\a^j}\tilde\rho{}^{ij}(\a)f_1(\a) \pa_{\a^i}f_2(\a)\,.
\]
This suggests to define the integral \equ{prod-two-b-insert} by
substituting $\pa_{\a^j}\tilde\rho{}^{ij}$ with
the weight function $\tilde\rho{}^{j}$, solution of 
\equ{div-tilde-rho} -- which has the correct transformation property to make the
integral invariant:
\eq 
\dint d^3\a \tilde\rho{}^{ij}(\a)\pa_{\a^i}f_1(\a) \pa_{\a^j}f_2(\a)
:= \half \dint d^3\a \tilde\rho{}^{j}(\a)
\lp f_1(\a) \pa_{\a^j}f_2(\a) - \pa_{\a^j}f_1(\a) f_2(\a) \rp\,.
\eqn{def-prod-two-b-insert} 
Recapitulating, the integration weights we shall need are the Haar measure $\rho_{\rm H}$
and the weight $\tilde\rho{}^{i}$ solution of \equ{div-tilde-rho}.

For the present gauge group SU(2) with the parametrization
\equ{g(alpha)}, these two functions explicitly read
\eq\ba{lll}
\rho_{\rm H}(\a) &=& \dfrac{4\sin^2(r/2)}{r^2} \,,\esp
\tilde\rho{}^{i}(\a) &=&   2  \dfrac{\a^i}{r^3}(r-\sin r) \,,
\ea\eqn{explicit-measures}
with $r^2=(\a^1)^2 + (\a^2)^2 + (\a^3)^2$.
We have imposed the normalization condition
$\int d^3\a d^3\b\,\rho(\a,\b) = 1$, which is non-trivial only 
for the first one.

\subsubsection{Internal product}\label{internal-product}

 A Hermitean internal product of two supergauge invariant 
state vectors $\Psi_1$ and $\Psi_2$
belonging to Cyl, given as
in \equ{wave-funct} with functions $\p_1$ and $\p_2$ of the holonomies $h[A]$
and $B\rhd h[A]$ along the space $S_1$ will now be defined with the help of the invariant
integration measure \equ{imt1} we have constructed:
\eq
\vev{\Psi_1|\Psi_2} = \dint_{\bf G} d\mu(\GG) \lp \p_1 (\GG)\rp^*  \p_2(\GG)
=\dint_{\bf G} d^3\alpha d^3\beta\,\rho(\alpha,\beta)
\lp F_1(\alpha,\beta)\rp^* F_2(\alpha,\beta) \,,
\eqn{int-prod}
where $^*$ means complex conjugation.
For an even vector $\Psi_+$, the corresponding function $F$ is equal to a
function $f_+(\trace g(\a))$, and for an odd vector, the corresponding
function is of the form
\eq
F(\a,\b) = Qf_-(\trace g(\a))= \b^i \pa_{\a^i} f_-(\trace g(\a))\,.
\eqn{odd-wave}
We have thus the following three internal product formula, obtained
using  \equ{integrals} and \equ{rho_3}, \equ{def-prod-two-b-insert}, 
and the notation $\chi=\trace g$:
\[\ba{l}
\vev{\Psi_{1+}|\Psi_{2+}} = \dint d^3\a\, \rho_{\rm H}(\a)
\lp f_{1+}(\chi(\a))\rp^* f_{2+}(\chi(\a))\,,\esp
\vev{\Psi_{1+}|\Psi_{2-}} = \dint d^3\a\, \tilde\rho{}^i(\a)
\lp f_{1+}(\chi(\a))\rp^* \pa_{\a^i} f_{2-}(\chi(\a))\,,\esp
\vev{\Psi_{1-}|\Psi_{2-}}
=i\dint d^3\a\, \tilde\rho{}^{ij}(\a)
\lp \pa_{\a^i} f_{1-}(\chi(\a))\rp^* \pa_{\a^j}f_{2-}(\chi(\a)) \esp
\quad = \dfrac{i}{2}\dint d^3\a\, \tilde\rho{}^j(\a)
\lp (f_{1-}(\chi(\a)))^* \pa_{\a^j} f_{2-}(\chi(\a)) - 
  (\pa_{\a^j}f_{1-}(\chi(\a)))^*  f_{2-}(\chi(\a)) \rp \,,
\ea\]
the last equality following from \equ{def-prod-two-b-insert}.

Using the supersymmetry singlet and doublet structure of the state
vector space, we can take as a basis the state vectors $\ket{j+}$ (even) and 
$\ket{j-}$ (odd) defined by
\eq
\vev{A|j+} = \trace R^{(j)}( h[A]) \,,\quad 
\vev{A,B|j-} =  Q\trace R^{(j)}(h[A])  \,,
\eqn{basis}
where $R^{(j)}$ is the spin $j$ representation matrix of $g$ $\in$ SU(2), 
$j$ = $0,\frac12,1,\cdots$. 
The even part of the basis is just the spin network basis of the loop quantization of the 
bosonic two-dimensional $BF$ model~\cite{Livine:2003kn}. Observe that the only supersymmetry 
singlet is the null spin state $\ket{0}$.

\subsubsection{Observables}

The quantization of the observables $W_0,\,W_1$, \equ{ott1}, and 
$L_0,\,L_1$, \equ{ott3}, is rather straightforward. The first two act
by multiplication (there are mapping cylindrical functions to cylindrical functions):
\[
\hat W_{0,1} \Psi[A,B] n= W_{0,1} \Psi[A,B]\,.
\]
Let us calculate the action of $L_0$ and $L_1$ on the basis vectors
\equ{basis}. When acting on wave functionals, they are represented by
the operators
\[
\hat L_0(x) = -\h^2 \dfud{}{A(x)}\cdot \dfud{}{B(x)}\,,\quad
\hat L_1(x) = -\h^2 \dfud{}{A(x)}\cdot \dfud{}{A(x)}\,,
\]
and the supersymmetry generator $Q$ by
\[
\hat Q = \dint dx B(x)\cdot \dfud{}{A(x)}\,.
\]
These operators obey the (anti-)commutation relations
\[
[\hat L_0,\hat Q]_+ = \hat L_1\,,\quad [\hat L_1,\hat Q]_- = 0\,.
\]
From these (anti-)commutation relations and the result~\cite{Livine:2003kn,rovelli}
\[
\dfud{}{A(x)}\cdot \dfud{}{A(x)} R^{(j)}( h[A]) 
= -j(j+1)R^{(j)}(h[A])\,,
\]
one easily shows that
\eq\ba{ll}
\hat L_0 \ket{j+}=0\,\quad &\hat L_0 \ket{j-}= \h^2j(j+1)\ket{j+}\,,\esp
\hat L_1 \ket{j+}= \h^2j(j+1)\ket{j+}\,,\quad 
&\hat L_1 \ket{j-}= \h^2j(j+1)\ket{j-}\,.
\ea\eqn{action-obs}
One sees that each supersymmetry doublet $\ket{j+}$, $\ket{j-}$ is
eigenvector of the bosonic observable $\hat L_1$ with the same
eigenvalue $\h^2j(j+1)$, whereas the fermionic observable $\hat L_0$ plays
the role of a step operator.

\section{Concluding remarks.}

After having recognized that the 
model of Ref.~\cite{schweda} had a rigid supersymmetry whose effect is to promote 
the full gauge symmetry of this model to a supergauge symmetry group ${\bf G}$,
the main step has been to construct an invariant measure of integration
on ${\bf G}$. This was achieved for a restricted class of
integrants, suitable for the purpose of defining an internal product.

We have thus obtained a full quantization of the $N=1$ supersymmetric
extension of the SU(2) $BF$-model in two-dimensions, with a basis of the 
physical space given by a supersymmetry singlet $\ket{0}$ and doublets
indexed by half-integer spin numbers $j\ge\frac12$. 
 We have also obtained explicit expressions for the action of the
observables on the state vectors.




\begin{thebibliography}{99}
\bibitem{Jackiw:1982hg}
R.~Jackiw.
\newblock {Liouville Field Theory: A Two-Dimensional Model for Gravity?}
\newblock In Bristol Adam~Hilgar Ltd, editor, {\em Quantum Theory of Gravity:
  Essays in Honor of the 60th Birthday of Bryce S. DeWitt}, pages 403--420.
  Christensen, S.M. ( Ed.), 1984.

\bibitem{Jackiw:1984je}
R.~Jackiw.
\newblock {Lower Dimensional Gravity}.
\newblock {\em Nucl.Phys.}, B252:343--356, 1985.

\bibitem{Jackiw:1995hb}
R.~Jackiw.
\newblock {Two Lectures on Two-Dimensional Gravity}.
\newblock {arXiv:gr-qc/9511048}, 1995.

\bibitem{Teitelboim:1983ux}  
C.~Teitelboim.
\newblock {Gravitation and Hamiltonian Structure in Two Space-Time Dimensions}.
\newblock {\em Phys.Lett.}, B126:41--45, 1983.

\bibitem{Teitelboim:1983fg}
C.~Teitelboim.
\newblock {The Hamiltonian Structure of Two-Dimensional Space-Time and its
  Relation with Conformal Anomaly}.
\newblock In Bristol Adam Hilger~Ltd, editor, {\em Quantum Theory of Gravity:
  Essays in Honor of the 60th Birthday of Bryce S. DeWitt}, pages 327--344.
  Christensen, S.m. ( Ed.), 1984.

\bibitem{Isler:1989hq} 
K. Isler and C. A. Trugenberger,
\newblock {A Gauge Theory of Two-Dimensional Quantum Gravity}.
\newblock {\em Phys.Rev.Lett.}, 63: 834-836, 1989.

\bibitem{Fukuyama:1985gg}   
Takeshi Fukuyama and Kiyoshi Kamimura, 
\newblock {Gauge Theory of Two-Dimensional Gravity}.
\newblock {\em Phys.Lett.}, B160: 259, 1985.

\bibitem{Livine:2003kn}   
E.R. Livine, Alejandro Perez, and C.~Rovelli.
\newblock {2D Manifold-Independent Spinfoam Theory}.
\newblock {\em Class.Quant.Grav.}, 63:835, 1989.

\bibitem{IvanMTese}
L.~Ivan~M. Bautista.
\newblock {Formalismo Hamiltoniano do Modelo de Jackiw-Teitelboim no Calibre
  temporal}, 2007, http://www.cce.ufes.br/pgfis/Disserta{\c
  c}{\~o}es/D-Luis

\bibitem{Constantinidis:2008br}   
Clisthenis~P. Constantinidis, J. Andre Lourenco, Ivan Morales, Olivier
  Piguet, and Alex Rios.
\newblock {Canonical Analysis of the Jackiw-Teitelboim Model in the Temporal
  Gauge. I. The Classical Theory}.
\newblock {\em Class.Quant.Grav.}, 25:125003, 2008.

\bibitem{Constantinidis:2008ty}
Clisthenis~P. Constantinidis, Olivier Piguet, and Alejandro Perez.
\newblock {Quantization of the Jackiw-Teitelboim Model}.
\newblock {\em Phys.Rev.}, D79:084007, 2009.

\bibitem{schweda} R.Leitgeb, M. Schweda, H.Zerrouki, \newblock {Finiteness of 2-D Topological $BF$ Theory with Matter Coupling}.
\newblock {\em Nucl.Phys.}, B542:425, 1999. arXiv:hep-th/9904204.

\bibitem{berezin}
F.A. Berezin, (Ed.) A.A. Kirillov and (Ed.) D. Leites,
\newblock {\em {Introduction to Superanalysis}}.
\newblock D Reidel Publishing Company, 1987.

\bibitem{corn} J.F. Cornwell.
\newblock {\em {Group Theory In Physics. VOL. III: Supersymmetries and Infinite
  Dimensional Algebras}}.
\newblock Academic Press, 1989.

\bibitem{henn}
M.~Henneaux and C.~Teitelboim.
\newblock {\em Quantization of Gauge Systems}.
\newblock Princeton University Press, 1992.

\bibitem{dirac} P. A. Dirac, \newblock {Lecture in Quantum Mechanics}.
\newblock Belfer Graduate School of Science, Yeshiva University, New York, 1964.

\bibitem{rovelli} Carlo Rovelli.
\newblock {\em {Quantum Gravity}}.
\newblock Cambridge Monographs on Mathematical Physics, 2004.

\end{thebibliography}
\end{document}